\documentclass[prb,aps,showpacs,twocolumn,amsmath,amssymb,floatfix,superscriptaddress]{revtex4-2}

\usepackage{hyperref}
\usepackage{graphicx}
\usepackage{bm}
\usepackage{xcolor}
\usepackage[utf8]{inputenc}
\usepackage{amssymb}
\usepackage{bbm}
\usepackage[normalem]{ulem}

\newcommand{\beq}{\begin{equation}}
\newcommand{\eeq}{\end{equation}}
\newcommand{\beqa}{\begin{eqnarray}}
\newcommand{\eeqa}{\end{eqnarray}}
\newcommand{\Tr}{\mathrm{Tr}}
\newcommand{\nn}{\nonumber}
\newcommand{\sect}[1]{\textit{#1.---}\ignorespaces}
\newcommand{\bi}{{\bm{i}}}
\newcommand{\bj}{{\bm{j}}}

\newcommand{\bl}{{\bm{l}}}

\newcommand{\editP}[1]{\textcolor{black}{#1}}

\begin{document}

\title{Exponential suppression of the topological gap in self-consistent intrinsic Majorana nanowires}

\author{Francisco Lobo}
\affiliation{Instituto de Ciencia de Materiales de Madrid (ICMM), CSIC, Madrid, Spain}
\author{Elsa Prada}
\affiliation{Instituto de Ciencia de Materiales de Madrid (ICMM), CSIC, Madrid, Spain}
\author{Pablo San-Jose}
\affiliation{Instituto de Ciencia de Materiales de Madrid (ICMM), CSIC, Madrid, Spain}

\date{\today}

\begin{abstract}
Predictions of topological p-wave superconductivity and Majorana zero modes (MZMs) in hybrid superconductor-semiconductor nanowires have been difficult to realize experimentally. Consequently, researchers are actively exploring alternative platforms for MZMs. In this work, we theoretically study depleted nanowires with intrinsic superconductivity (as opposed to proximity-induced). Using a self-consistent Hartree-Fock-Bogoliubov mean field theory, we compute the topological phase diagram versus Zeeman field and filling for intrinsic wires with attractive interactions. We find that, although intrinsic wires could be less vulnerable than hybrids to topology-adverse effects, such as disorder and metallization, they are hindered by a fundamental limitation of their own.  Although a topological p-wave gap is indeed possible, it is far less robust than in hybrid Majorana nanowires. Instead of remaining stable beyond the topological transition, it is found to decay exponentially with Zeema field, greatly reducing the parameter region with an appreciable topological gap.
\end{abstract}

\maketitle

Among the known emergent states associated with topological materials, Majorana zero modes (MZMs) in topological p-wave superconductors occupy a prominent place~\cite{Kitaev:PU01,Alicea:RPP12,Aguado:RNC17}. The properties of MZMs are remarkable even compared to other kinds of topologically protected states: they are spatially bound, zero-charge, zero-spin and zero-energy fractionalized quasiparticles, immune to local perturbations, and with non-Abelian anyon statistics~\cite{Lutchyn:NRM18,Beenakker:ARCM13,Alicea:RPP12}. They are thus able to form a decoherence-free subspace to store and geometrically transform quantum information through the braiding group~\cite{Nayak:RMP08}. Unfortunately, and because of their very nature, MZMs have proven to be elusive. Despite our best continued efforts and the many potential sightings in experiments ~\cite{Mourik:S12,Gul:NN18,Vaitiekenas:S20,Deng:S16,Deng:PRB18,Schiela:PQ24,Aghaee:PRB23}, candidate MZMs refuse to behave as predicted by theory, so considerable questions still remain about these detections ~\cite{Prada:NRP20,Yu:NP21,Valentini:S21,Valentini:N22,Pan:PRR20,Kouwenhoven:MPLB24}.

The most studied platform for MZMs is known as the Majorana hybrid nanowire~\cite{Oreg:PRL10,Lutchyn:PRL10}. It is composed of a semiconductor nanowire with strong spin-orbit coupling (SOC) partially or fully covered by an s-wave superconductor. Under a strong Zeeman field above a critical value, this kind of device is predicted to undergo a transition into a one-dimensional (1D) p-wave topological superconductor phase with MZMs at its ends. This hybrid wire approach to MZMs has spurred great interest and excitement in the last decade because of its elegance, experimental feasibility, and apparent simplicity. The simplicity was, unfortunately, deceptive. Even the best hybrid devices were riddled with a host of unwanted effects~\cite{Prada:NRP20}, notably metallization~\cite{Reeg:PRB18}, disorder~\cite{Pan:PRR20,Ahn:PRM21,Das-Sarma:NP23}, trivial state pinning~\cite{Dominguez:NQM17,Liu:PRB17c}, smooth confinement~\cite{Kells:PRB12,Prada:PRB12,Penaranda:PRB18,Stanescu:PRB19,Vuik:SP19}, etc., which introduced the possibility of various types of false positives. In view of these problems, some researchers have slowly turned their efforts towards other possible realizations of p-wave superconductivity. One minimal variation is to make the Majorana nanowire ``intrinsic" (as opposed to hybrid) by fabricating it from a \editP{clean semiconductor material with strong SOC that possesses a superconducting phase even at low carrier densities. Possible examples include LaAlO${}_3$/SrTiO${}_3$ oxide interfaces~\cite{Perroni:PRB19,Barthelemy:EL21}. Also 2D materials such as bilayer graphene/WS${}_2$ heterostructures~\cite{Zhang:N23,Zhou:S22} or multilayer MoS${}_2$ crystals~\cite{Lu:S15a}, which have recently been demonstrated to develop low-density superconducting phases~\cite{Crepel:PNAS22}. Since superconductivity would be intrinsic to a nanowire crafted from these materials, the issues associated with the contact to} an electronically dense s-wave superconductor would be avoided. The predictions of D-class~\cite{Schnyder:PRB08} topological phases valid for hybrid nanowires do not necessarily apply to the intrinsic case. The conventional hybrid-wire description with externally fixed order parameter becomes inaccurate, and a more precise, self-consistent characterization of the intrinsic pairing potential as a function of magnetic field and filling becomes essential.

\begin{figure}
    \includegraphics[width=\columnwidth]{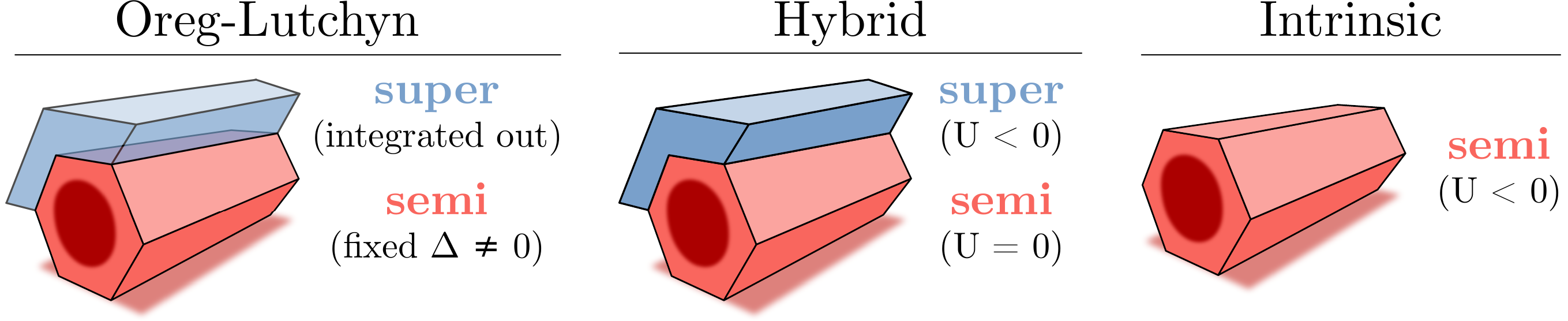}
    \caption{Three models for Majorana nanowires. $U$ is an onsite Hubbard interaction, $\Delta$ is a superconducting pairing \editP{induced by proximity to a superconducting shell (blue region). All these models are 1D or quasi-1D approximations, valid for sufficiently depleted semiconductors (red region).}}
    \label{fig:1}
\end{figure}

In this work, we study the topological phase diagram of an intrinsic Majorana nanowire within a self-consistent Hartree-Fock-Bogoliubov (HFB) mean-field theory of interactions~\cite{El-Bassem:NPA17,Younes:19} and compare it to its hybrid counterpart. We demonstrate that while a p-wave phase is indeed possible in the intrinsic case, it is far less robust than in the hybrid Majorana nanowire. Instead of remaining stable beyond the critical magnetic field, it is found to decay exponentially with field, greatly reducing the region in the phase diagram with an appreciable topological gap.
This behavior also destroys Majorana oscillations with magnetic field and Fermi energy, which are considered one of the most characteristic distinguishing features of MZMs in hybrid nanowires~\cite{Das-Sarma:PRB12}.
We additionally show that long-range attractive interactions contribute to a slower, but still exponential decay of the topological gap while also substantially changing the topological phase boundary.

\sect{Self-consistent superconductivity}
We start by presenting a summary of the self-consistent Bardeen-Cooper-Schriefer (BCS) and HFB theories of superconductivity applied to the attractive Hubbard model. Full details are presented in the Appendices.

Consider a Hubbard model on a lattice with attractive onsite interactions $U<0$,
\beqa
\label{Hubbard}
H^\mathrm{Hub} &=& H_0+H_U,\\
H_0 &=& -t\sum_{\langle i,j\rangle,\sigma} c_{i\sigma}^\dagger c_{j\sigma}  + (2t-\mu)\sum_{i,\sigma} n_{i\sigma},\\
H_U &=& U\sum_i n_{i\uparrow}n_{i\downarrow}.
\label{HU}
\eeqa
In the above equation, $\langle i,j\rangle$ denotes nearest-neighbor site indices, $\mu$ the Fermi energy, $t = \hbar^2/2ma_0^2$ the hopping amplitude, $m$ the effective mass, $a_0$ the lattice spacing, $c_{i\sigma}^\dagger$ ($c_{i\sigma}$) the creation (destruction) particle operators at site $i$ with spin $\sigma=\uparrow,\downarrow$, and $n_{i\sigma} = c_{i\sigma}^\dagger c_{i\sigma}$ the number operator.


The BCS theory of superconductivity in metals~\cite{Bardeen:PR57} predicts that, below a certain critical temperature $T_c$, the attractive ($U<0$) interaction will give rise to a finite complex-valued superconducting order parameter $\Delta^{ii} = U \langle c_{i\uparrow}  c_{i\downarrow}\rangle$. The anomalous order parameter $\Delta^{ii}$ creates a superconducting gap $\Omega$ for quasiparticle excitations that corresponds to the binding energy of Cooper pairs in the BCS condensate. (In a uniform and infinite $U<0$ Hubbard model with time-reversal symmetry, $\Omega=|\Delta|$, but not necessarily in more general situations.) The key insight of the BCS theory is that by relaxing the constraint of a fixed number of particles~\cite{Bardeen:PR57,Tinkham:04}, $U<0$ will naturally lead, by spontaneous breaking of the gauge symmetry, to the emergence of a non-zero $\Delta^{ii}$ at a self-consistent mean-field level, transforming Eq. \eqref{Hubbard} into $H^\mathrm{Hub}_\mathrm{BCS} = H_0 + \Sigma_\mathrm{BCS}$, where
\beq
\Sigma_\mathrm{BCS} =
\sum_{i} \Delta^{ii} c^\dagger_{i\downarrow} c^\dagger_{i\uparrow}+\text{h.c.}
\label{SigmaBCS}
\eeq
This $\Sigma_\mathrm{BCS}$ is the mean-field version of the interaction term $H_U$.
The self-consistent mean-field condition for $\Delta^{ii}$ is straightforward to obtain in the case of an infinite, uniform system (see Appendix \ref{ap:HFB}), and is given by the so-called ``gap equation'',
\beq
\label{gapeq0main}
\Delta = - \frac{U}{N}\sum_k\frac{\Delta}{2E_k}\tanh\left(\frac{E_k}{2k_BT}\right),
\eeq
where $\Delta = \Delta^{ii}$ (assumed real without loss of generality), $N$ is the number of sites, $T$ is temperature, $E_k = \sqrt{\Delta^2+\epsilon_k^2}$ and $\epsilon_k$ are the eigenvalues of $H_0$ and $\Delta$. The above gap equation is valid only in the Andreev limit of dense metals, that is, as long as
\beq
\label{Andreev}
\Delta\ll \mu.
\eeq
It also assumes no spin-mixing terms in $H_0$.

Although the above form of BCS theory is valid in conventional superconductors, it is not applicable to Majorana nanowires, nearly-depleted semiconductors with strong SOC and magnetic fields. The correct mean-field treatment of superconductivity in this case requires a generalization of the gap equation, known as HFB theory.

The HFB theory does not make any assumption about the normal Hamiltonian $H_0$ or the Fermi energy, and therefore needs to keep track of all the possible local order parameters, each of which influences the others. These can all be written in terms of different components of the local density matrix in Nambu space, including both its normal electron-electron ($ee$) and ``anomalous'' hole-electron ($he$) blocks,
\beq
\rho^{ii}_{e\sigma,e\sigma'} = \langle c^\dagger_{i\sigma'}c_{i\sigma}\rangle, \medspace\medspace\medspace \rho^{ii}_{h\sigma,e\sigma'} = \langle c_{i\sigma'}c_{i\sigma}\rangle.
\label{rho_simple}
\eeq

These can be computed in equilibrium by summing all thermally occupied Bogoliubov eigenstantes of the complete mean-field Hamiltonian $H_\mathrm{HFB}=H_0+\Sigma_\mathrm{HFB}$, where the self-energy $\Sigma_\mathrm{BCS}$ in Eq. \eqref{SigmaBCS} has been replaced by the more general $\Sigma_\mathrm{HFB}$ that reads
\beqa
\label{HFB}
\Sigma_\mathrm{HFB} &=& \frac{1}{2}\sum_{i\sigma\sigma'}(c^\dagger_{i\sigma},c_{i\sigma})\check{\Sigma}_{\sigma\sigma'}^{ii}
\left(\begin{array}{c}
c_{i\sigma'}\\
c^\dagger_{i\sigma'}
\end{array}\right),\\
\label{SigmaHFB}
\check{\Sigma}^{ij} &=& U\delta_{ij}\left(\frac{1}{2}\mathrm{Tr}(\tau_z \tilde{\rho}^{ii})\tau_z - \tau_z\tilde{\rho}^{ii}\tau_z\right),\\
\label{checkrho}
\tilde{\rho}^{ij} &=& \left( \begin{array}{cc}
    \rho_{ee}^{ij} & (\rho_{he}^{ji})^\dagger \\
    \rho_{he}^{ij} & -(\rho_{ee}^{ji})^T
\end{array} \right).
\eeqa
Here $\tau_z$ is the third Pauli matrix on the Nambu (electron/hole) space and the trace is taken over both Nambu and spin spaces [we have omitted the spin indices in Eqs. \eqref{SigmaHFB} and \eqref{checkrho} for brevity]. The above equation, together with Eq. \eqref{rho_simple}, define a set of self-consistent equations with $2\times 2$ matrices $\rho^{ii}_{ee}$ and $\rho^{ii}_{he}$ as unknowns. To solve them, one typically starts from a seed guess and then uses the $H_\mathrm{HFB}$ they define to obtain a new density matrix, iterating until convergence.

The expression for $\Sigma_\mathrm{HFB}$ is valid beyond the standard BCS theory, for arbitrary filling, SOC and Zeeman. Appendix \ref{ap:HFB} presents a compact derivation, also generalized to interactions of arbitrary range. In Appendix \ref{ap:HFBnosoc} we derive, in the case of $\alpha=0$, an analytic generalization of the self-consistent gap equation for $\Delta$, Eq. \eqref{gapeq0main}, and also for the renormalized $\mu$ and the Zeeman field.

\begin{figure}
\includegraphics[width=0.95\columnwidth]{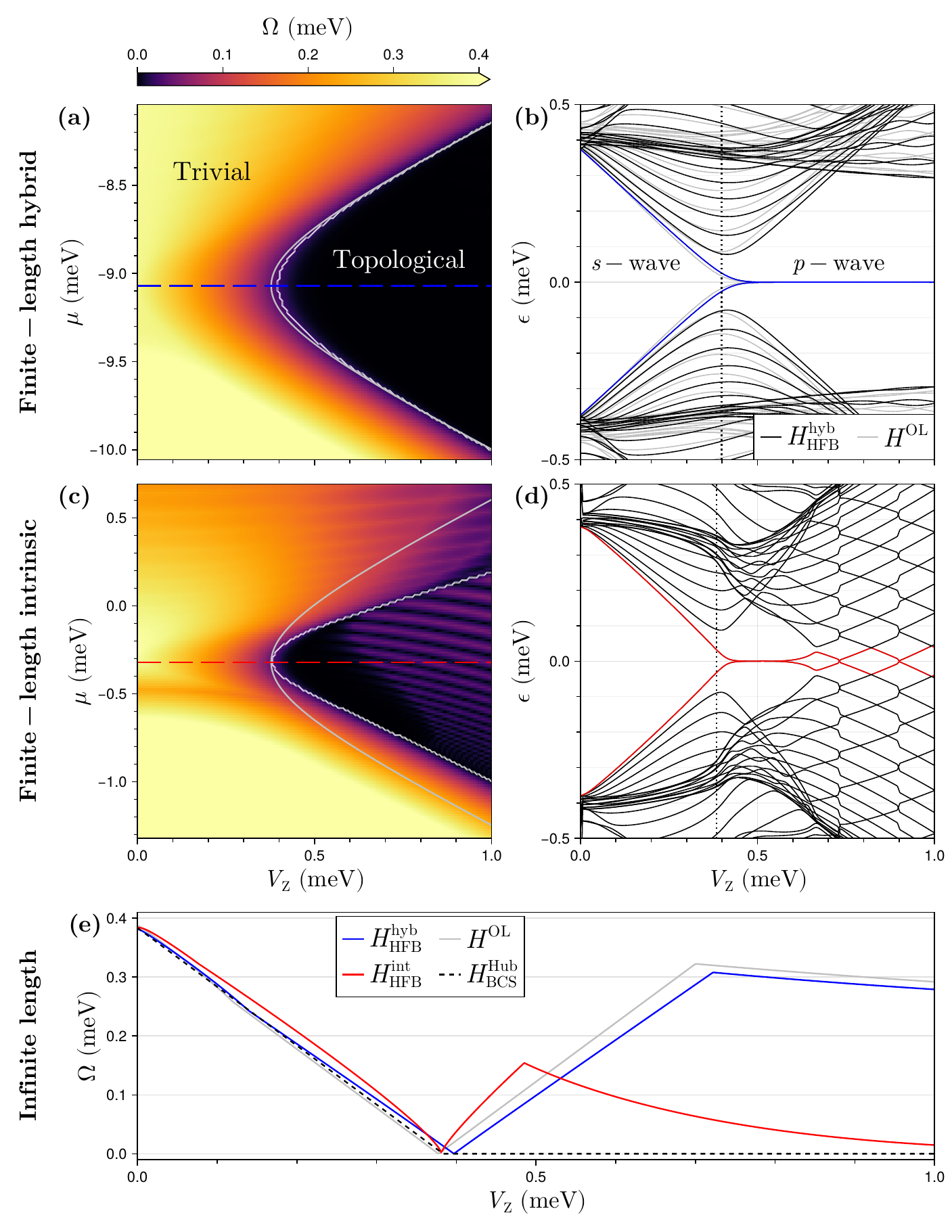}
    \caption{(a) Topological phase diagram of the finite-length, self-consistent hybrid Majorana nanowire model as a function of Zeeman field $V_Z$ and Fermi energy $\mu$ in the semiconductor. Color encodes the energy $\Omega$ of the lowest-lying excitation at each point. The white line marks the topological transition in bulk, and the gray line the transition in the Oreg-Lutchyn (OL) model for comparison. (b) Linecut of the low-energy spectrum along the blue dashed line in (a), exhibiting a topological transition and the emergence of non-oscillating Majorana zero modes (MZMs) beyond a critical Zeeman field $V_Z^c$. The spectrum of the OL model is shown in gray. (c,d) Analogues of (a,b) for the self-consistent intrinsic nanowire model. (e) $\Omega$ as a function of $V_Z$ in an infinite nanowire for all three models. \editP{Since it is infinite, any near-zero energy levels from boundary states are absent in the spectrum, and $\Omega$ for $V_Z>V_Z^c$ measures the topological minigap in the bulk.} The dashed line is the analytical result from the generalized gap equation without SOC, see Appendix \ref{ap:HFBnosoc}, showing that SOC is essential to develop a topological p-wave gap in Majorana nanowires. Hybrid parameters:  $U=-32$meV, $\mu_{\rm{SC}}=10$meV, semiconductor g-factor $g=12$ ($g_{s}=2$ in the superconductor), SC-SM hopping $t'=0.8t$. Intrinsic parameters: $U=-8.0$meV. Common parameters: wire length $L=2000$nm, lattice constant $a_{0}=10$nm, effective mass $m=0.023m_{e}$, SOC $\alpha=40$meV.}
    \label{fig:2}
\end{figure}

The HFB theory allows one to explore self-consistent superconductivity in Majorana nanowires, where the standard BCS assumptions are violated in three ways: the wires are close to depletion, they include SOC-induced spin-mixing, and they are subject to Zeeman fields. In the following, we analyze and compare, in the context of HFB theory, the phase diagram of three models relevant to Majorana nanowires: the Oreg-Lutchyn model, the self-consistent hybrid model and the self-consistent intrinsic model, see Fig. \ref{fig:1}.

All the numerical calculations that follow are performed using the Quantica.jl  Julia package \cite{San-Jose:24}. Self-consistent mean field iteration is performed using Anderson mixing, and density matrices are computed using exact diagonalization (finite wires) or adaptive k-space integration (infinite wires).


\sect{The Oreg-Lutchyn model} The Oreg-Lutchyn (OL) model~\cite{Oreg:PRL10,Lutchyn:PRL10} has been extensively used to describe hybrid Majorana nanowires, a type of semiconductor nanowire (typically made of InAs or InSb) on top of which a conventional superconductor shell (typically made of Al) is deposited or grown. The nanowire has strong Rashba SOC $\alpha k_z\sigma_y$ ($k_z$ is the electron wavevector along the wire) and is subject to an external Zeeman field $V_Z \sigma_z= \frac{1}{2}g B_z\sigma_z$, where $g$ is the g-factor, $B_z$ the applied magnetic field, and $\sigma$ are spin Pauli matrices. The shell induces superconductivity on the semiconductor by proximity, which the model treats at the simplest possible level, by adding a constant BCS pairing to the semiconductor Hamiltonian
\beqa
H^\mathrm{OL} &=& H_0+H_{\rm{SOC}}+H_Z+\Sigma_\mathrm{BCS},\\
H_{\rm{SOC}} &=& \frac{\alpha}{2a_0}\sum_{i,\sigma} \left(c_{i+1\bar\sigma}^\dagger c_{i,\sigma} + \text{h.c.}\right),\\
H_{Z} &=& V_Z\sum_{i}  \left(c_{i\uparrow}^\dagger c_{i\uparrow}-c_{i\downarrow}^\dagger c_{i\downarrow}\right).
\eeqa
The spin $\bar\sigma$ denotes the opposite of $\sigma$. The pairing $\Delta$ is fixed to a certain value related to the shell pairing and the transparency of the contact. Thus, any self-consistent dependence of $\Delta$ on other parameters or on position is ignored.

The basic properties of the OL model have been studied in great detail~\cite{Alicea:RPP12,Marra:JAP22}. Here we summarize three key results: (i) The infinite OL model undergoes a band inversion as $V_Z$ exceeds a critical value $V_Z^c=\sqrt{\Delta^2+\mu^2}$. The system then develops a topological p-wave superconducting phase with a mini-gap that grows with spin-orbit strength. (ii) For long but finite nanowires with $V_Z>V_Z^c$, zero-energy Majorana bound states appear at the ends of the wire, whose wave function extent $L_M$ is inversely proportional to the topological minigap. (iii) When the length $L$ of the topological nanowire decreases below $2L_M$, the two end Majorana bound states begin to overlap and split, oscillating around zero energy as $\pm\delta\epsilon\sim\pm e^{-2L/L_M}\cos(k_F L)$, where $k_F$ is the Fermi wavevector in the normal phase.

We show in the following that this model and its predictions are qualitatively accurate compared to a self-consistent hybrid nanowire, as long as the parent superconductor satisfies Eq. \eqref{Andreev}. However, it fails to describe self-consistent intrinsic Majorana nanowires, where superconductivity comes from attractive interactions in the semiconductor wire itself.


\sect{The self-consistent hybrid nanowire model} The hybrid model Hamiltonian includes the electrons of the shell superconductor explicitly in order to compute its superconductivity self-consistently as a function of parameters,
\beqa
H^\mathrm{hyb} &=& H^{\rm{SM}}_0+H^{\rm{SM}}_\mathrm{SOC} + H^{\rm{SM}}_\mathrm{Z}+ \\
&& H^{\rm{SC}}_0+H^{\rm{SC}}_U + H^{\rm{SC}}_\mathrm{Z} + H^{\rm{SC-SM}},\nn\\
H^\mathrm{SC-SM} &=&- t' \sum^{\substack{i\in \rm{SC} \\ j\in \rm{SM}}}_{\langle i,j\rangle\sigma} c_{i\sigma}^\dagger c_{j\sigma} + \text{h.c.}
\eeqa




The labels SM and SC indicate two distinct parallel one-dimensional lattices, one for the semiconductor and the other for the superconductor. The coupling between them, $H^{\rm{SC-SM}}$, is governed by the hopping parameter $t'$. We assume that the Fermi energy in the superconductor (dubbed $\mu_{\rm{SC}}$, as opposed to $\mu$ in the semiconductor)
is large, so that the standard BCS theory could be applied to it in isolation, as done in Eq. \eqref{SigmaBCS}.

The HFB self-consistent mean-field approximation $H^\mathrm{hyb}_\mathrm{HFB}$ of the hybrid system as a whole can be obtained by replacing $H^\mathrm{SC}_U$ above with $\Sigma^\mathrm{SC}_\mathrm{HFB}$ following Eq. \eqref{SigmaHFB}, and solving for $\rho^{ii}_{ee}$ and $\rho^{ii}_{he}$ with Eq. \eqref{rho_simple}. This can be done either for an infinite and uniform nanowire (using the Bloch theorem to obtain position-independent $\rho$'s) or for a finite nanowire (so that $\rho^{ii}$ is site dependent). The gap $\Omega$ (that is, the smallest positive eigenvalue) of the resulting $H^\mathrm{hyb}_\mathrm{HFB}$ in the long but finite nanowire is shown in Fig. \ref{fig:2}(a) as a function of $V_Z$ and the semiconductor's $\mu$. The onsite interaction $U$ is chosen so that $\Omega(V_Z=0) = \Delta \approx 0.38$ meV. We see that $\Omega$ vanishes for $V_Z$ greater than $V_Z^c \approx \sqrt{\mu^2 + \Delta^2}$ (gray curve), as expected from the OL model, due to the emergence of a MZM at the boundaries. Figure \ref{fig:2}(b) shows the full spectrum along a cut with fixed $\mu$ (dashed blue line in Fig. \ref{fig:2}(a)). Since the nanowire in this simulation has $L>2L_M$, the smallest eigenvalues (solid blue line in Fig. \ref{fig:2}(b)) go to zero in the topological region, without any Majorana oscillations. For comparison, the non-self-consistent $H^\mathrm{OL}$ spectrum is shown in gray in Fig. \ref{fig:2}(b), matching with good accuracy the $H^\mathrm{hyb}_\mathrm{HFB}$ spectrum. As in the OL model, the self-consistent p-wave mini-gap above the zero mode is robust and remains finite as $V_Z$ increases, moving deeper into the topological region. Its evolution along the fixed-$\mu$ cut, computed in an infinite nanowire, is shown in blue in Fig. \ref{fig:2}(e), with the OL equivalent in light gray. \footnote{\editP{Note that for the infinite nanowire no boundary modes are present, so the $\Omega$ for $V_Z>V_Z^c$ in Fig. \ref{fig:2}(e) measures the minigap, not the MZM splitting.}} The qualitative agreement between the two models is the result of the parent superconductor satisfying the Andreev condition $\Delta\ll\mu_{\rm{SC}}$.

\begin{figure}
    \includegraphics[width=\columnwidth]{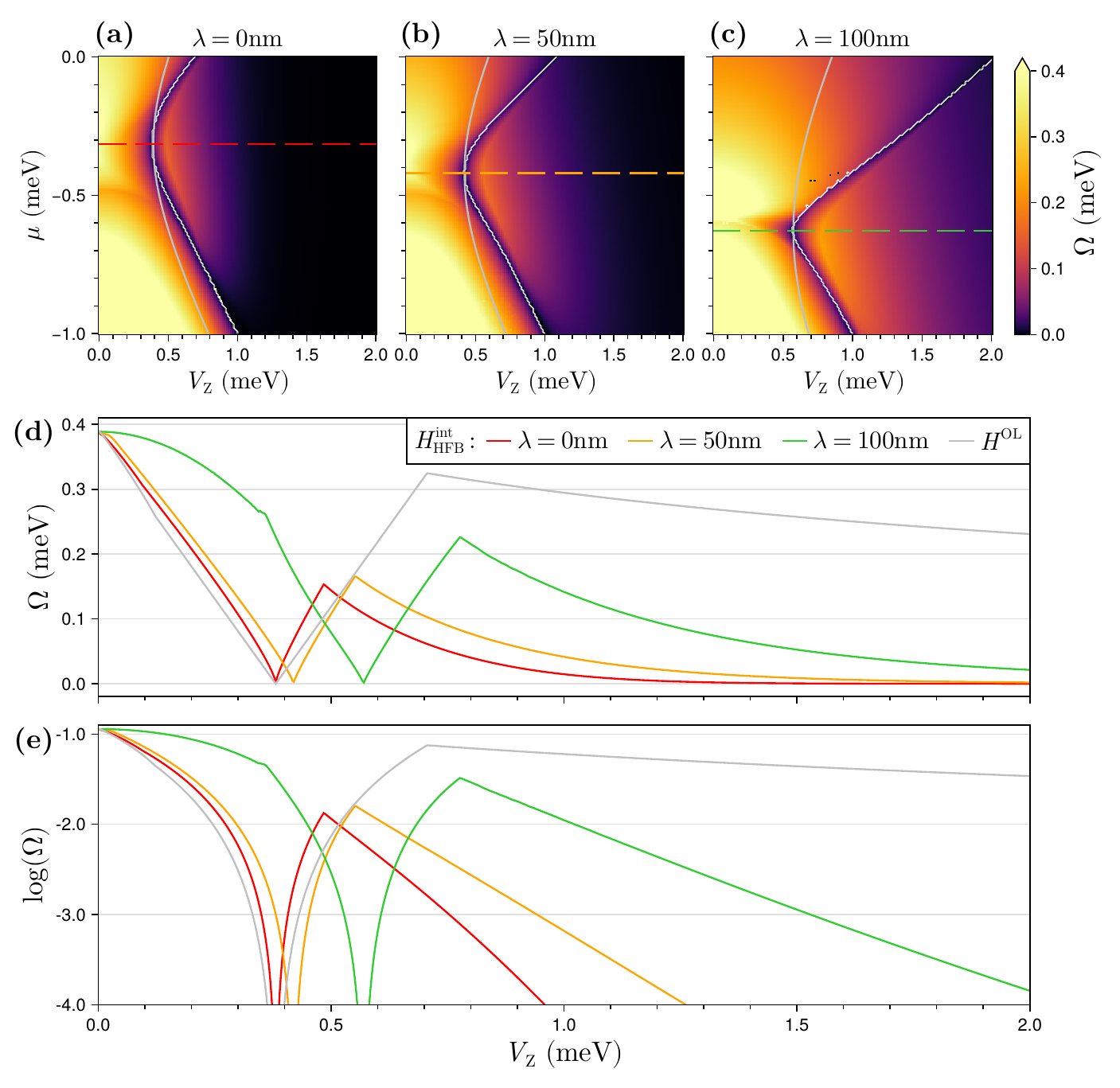}
    \caption{(a-c) Phase diagram of an infinite intrinsic nanowire with screened Coulomb interactions of increasing range, governed by the screening length $\lambda$.  The $\lambda = 0$ case corresponds to purely onsite interactions, see Fig. \ref{fig:2}. As $\lambda$ increases the topological boundary becomes sharper at its $V_Z^c$ minimum. $U$ is chosen in each case to yield a fixed $\Delta = 0.38$ meV at $V_Z=0$, namely $U=-8.0$meV, $U=-3.9$meV and $U=-4.3$meV, respectively (other parameters as in Fig. \ref{fig:2}). (d,e) The gaps along the horizontal cuts [dashed in (a-c)] become longer range, but are still exponentially decaying with $V_Z$, as shown by the log-plot (e). For comparison, the OL gap (gray) decays much more slowly, as a power-law.}
    \label{fig:3}
\end{figure}

\sect{The intrinsic nanowire model} The case of a Majorana nanowire with intrinsic superconductivity (i.e. not induced externally by a superconducting shell) is surprisingly different. The model for such a nanowire is defined by
\beq
H^{int} = H_0+H_\mathrm{SOC} + H_\mathrm{Z} + H_U.
\label{intrinsic}
\eeq
Note that attractive interactions $U<0$ are now present in the semiconductor itself.
Since the filling of the wire may be small, or even zero, the Andreev limit in Eq. \eqref{Andreev} is not satisfied. Moreover, the SOC term and Zeeman fields are no longer irrelevant to the superconducting order parameter. Therefore, it should not be expected that the mean field $H^{int}_\mathrm{HFB}$ necessarily matches the phenomenology of the OL model. The gap and spectrum of $H^{int}_\mathrm{HFB}$ are shown in Figs. \ref{fig:2}(c,d) for a finite wire with equivalent parameters as the hybrid case of Figs. \ref{fig:2}(a,b). We see that, indeed, while the topological boundary and band inversion are qualitatively similar, the behavior of the minigap in the topological region has a very different behavior. The minigap is dramatically more fragile, decreasing exponentially with $V_Z$ after the transition (red curve in Fig. \ref{fig:2}(e)), which leads to its collapse shortly after, instead of remaining roughly constant as in the hybrid case. As a result, the parameter space region with MZMs is strongly reduced to a narrow black strip in Fig. \ref{fig:2}(c). After the collapse of the minigap, the Majoranas quickly become delocalized finite-energy Andreev states. This is not the result of a suboptimal choice of model parameters but stems from the intrinsic origin of the p-wave pairing.

We confirmed that this type of behavior is also generic in the case of multimode intrinsic nanowires. We extend the nanowire model to have $M$ transverse modes by employing a quasi-1D lattice with $M$ sites in the transverse direction, generalizing also $H_{\rm{SOC}}$ to include an additional $\alpha k_y \sigma_z$ term. The filled subbands develop a finite pairing because of the attractive interactions. The shallowest mode can undergo a topological transition like for $M=1$. However, its p-wave minigap is not affected by the pairing of the rest of the orthogonal eigenmodes, so that the minigap still collapses shortly after the critical Zeeman $V^c_Z$ (not shown).

We have explored a second generalization: intrinsic nanowires with finite-range interactions. We replace $H_U$ in Eq. \eqref{intrinsic} with Eq. \eqref{Hint0} and $v^{ij} = \frac{1}{2} U a_0 e^{-|r_i-r_j|/\lambda}/|r_i-r_j|$ for $i\neq j$ (screened Coulomb interaction with screening length $\lambda$), while still imposing $v^{ii} = U$ on each site. The self-consistent HFB decoupling then results in a topological boundary that considerably deviates from the OL-like hyperbolic shape as $\lambda$ grows; see Fig. \eqref{fig:3}(a-c). The topological minigap becomes more robust, remaining finite farther into the topological region. However, it still decays exponentially, as shown in Figs. \eqref{fig:3}(d,e)

\sect{Conclusion}
The exponential decay of the minigap in intrinsic nanowires stems from competing requirements for the topological phase: spinlessness and superconductivity. As the Zeeman field $V_Z$ exceeds the critical $V_Z^c$, one spin sector is depleted. Thus, the probability of finding two electrons at a given site forming a singlet to condense into a Cooper pair vanishes in the absence of SOC (black dashed line in Fig. \ref{fig:2}(e)). With a finite SOC, this probability grows as a result of the SOC-induced spin canting of states with opposite momenta. The canting, and hence the singlet probability, decreases with $V_Z$. Since the mean-field self-consistent pairing depends exponentially on this probability, see Eq. \eqref{gapsol0}, we obtain an exponential decay of the pairing with $V_Z$. This is the underlying mechanism behind the fragile topology observed in our numerical simulations. In stark contrast, the pairing in the superconductor shell of a hybrid nanowire remains essentially unaffected by Zeeman, since its large Fermi energy guarantees spinful carriers and a large $V_Z$-independent singlet probability on any site. The proximity-induced pairing on the semiconductor then depends only linearly on the spin canting itself, which falls as a weak power law with $V_Z$.

Our results emphasize the fact that the pursuit of topological intrinsic superconductivity is hindered by a weak spin canting of spinless carriers. This limitation does not affect Majoranas obtained by the time-reversal-symmetric Fu-Kane approach \cite{Fu:PRL08}, but it is expected to make the Zeeman-driven OL mechanism problematic as a route toward intrinsic topological superconductivity.

\acknowledgements
We thank R. E. F. Silva and B. Amorim for enlightening discussions. This work has been supported by MICIU/AEI/10.13039/501100011033, `ERDF A way of making Europe' and `ESF+' through Grant PID2021-122769NB-I00.

\bibliography{biblio}
\appendix

\section{Hartree-Fock-Bogoliubov theory}
\label{ap:HFB}

Take a Hamiltonian with interactions of the form
\beqa
\label{H}
H &=& H_0 + H_{\text{int}}, \\
H_0 &=& \sum_{ij} c_i^\dagger H_0^{ij}c_j,\nn\\
H_{\text{int}} &=& \frac{1}{2} \sum_{ij}  c_i^\dagger c_i v^{ij} c_j^\dagger c_j - \frac{1}{2}\sum_i c_i^\dagger c_i v^{ii},
\label{Hint0}
\eeqa
where the last term removes spurious self-interactions.
Here we are assuming a spinless, single-orbital model, so that $i,j$ are purely spatial indices and
$v^{ij} = V(\mathbf{r}_i-\mathbf{r}_j)$ is a scalar potential between sites $i$ and $j$. We assume $v^{ij} = v^{ji}$ from this point on.
We may add extra orbital indices  $s_i$ at each site $i$ (such as spin), so that our interaction model becomes
\beqa
H_{\text{int}} &=& \frac{1}{2} \sum_{is_ijs_j}  c_{is_i}^\dagger c_{is_i} v^{ij} c_{js_j}^\dagger c_{js_j} - \frac{1}{2}\sum_{is_i} c_{is_i}^\dagger c_{is_i} v^{ii} \nn\\
&=& \frac{1}{2} \sum_{\bi\bj}  c_\bi^\dagger c_\bi v^{ij} c_\bj^\dagger c_\bj - \frac{1}{2}\sum_\bi c_\bi^\dagger c_\bi v^{ii}.
\eeqa
Here we have denoted composite index pairs, such as $(i,s_i)$, with bold indices, such as $\bi$. The interaction potential $v^{ij}$ is assumed independent of $s_i, s_j$.

A perturbation theory treatment of the interaction based on path integrals starts by casting the Hamiltonian into its normal-ordered form \footnote{Normal ordering of $H(c^\dagger,c)$ is required so that $\langle\phi_1|H(c^\dagger,c)|\phi_2\rangle = H(\bar\phi_1,\phi_2)\langle\phi_1|\phi_2\rangle$ for coherent states $|\phi_i\rangle$ and Grassmann variable $\phi_i$.}. Normal ordering cancels the self-interaction term in $H_{\text{int}}$,
\beq
H_{\text{int}} = \frac{1}{2} \sum_{\bi\bj}  c_\bi^\dagger c_\bj^\dagger v^{ij}  c_\bj c_\bi.
\eeq
The mean-field Hartree-Fock-Bogoliubov decoupling (i.e. including pairing) of the Hamiltonian then yields
\beqa
\label{meanfield}
H &\approx& H_0 + H_{\text{int}}^{H}+H_{\text{int}}^{F} + H_{\text{int}}^\mathrm{B}\\
&=& \sum_{\bi\bj}\left[c_\bi^\dagger \left(H_0^{\bi\bj}+\Sigma_{H}^{\bi\bj}+\Sigma_{F}^{\bi\bj}\right) c_\bj + \frac{1}{2}\left(c_\bi^\dagger \Sigma^{\bi\bj}_B c_\bj^\dagger +\mathrm{h.c.}\right)\right],\nonumber
\eeqa
where the Hartree (electrostatic), Fock (exchange) and Bogoliubov (pairing) self-energies between spatial sites $i$ and $j$ are matrices over orbital space given by \footnote{These self-energies stem from the Wick theorem over Grassmann variables when expanding $\exp(iS_{\text{int}}) = \exp[-i\int_C dt H_{\text{int}}(\bar\psi^\sigma,\psi^\sigma)]$ in the path integral. Since $H_{\text{int}}$ on each branch $\sigma=\pm$ of the contour $C$ involves only same-branch and equal-time fields $\psi^\sigma$, the self-energy ends up being proportional to $\langle \bar\psi^\sigma_j(t)\psi^\sigma_i(t)\rangle$, which is the density matrix. With retarded interactions, the self-energy would also involve retarded Green's functions of different times.}
\beqa
\Sigma_H^{ij} &=&\delta_{ij} \mathbbm{1} \sum_{\bl} v^{il}\langle c_{\bl}^\dagger c_{\bl}\rangle =
\delta_{ij} \mathbbm{1} \sum_{l} v^{il} \mathrm{Tr}\rho_{ee}^{ll},
\nn\\
\Sigma_F^{ij} &=& -v^{ij}\langle c^\dagger_\bj \otimes c_\bi\rangle =
-v^{ij}\rho_{ee}^{ij},
\nn\\
\Sigma_B^{ij} &=& v^{ij}\langle c_\bj \otimes c_\bi\rangle =
v^{ij}\rho_{eh}^{ij}.
\label{Sigma}
\eeqa
The electron-electron, hole-hole and the anomalous density matrices are defined as
\beqa
\rho_{ee}^{ij} &=& \langle c_\bj^\dagger \otimes c_\bi\rangle = (\rho_{ee}^{ji})^\dagger, \nn \\
\rho_{hh}^{ij} &=& \langle c_\bj \otimes c_\bi^\dagger\rangle =  (\rho_{hh}^{ji})^\dagger = \mathbbm{1}\delta_{ij} -  (\rho_{ee}^{ji})^T,\nn\\
\rho_{eh}^{ij} &=& \langle c_\bj \otimes c_\bi\rangle  = -(\rho_{eh}^{ji})^T = (\rho_{he}^{ji})^\dagger,\nn\\
\rho_{he}^{ij} &=& \langle c_\bj^\dagger \otimes c_\bi^\dagger\rangle = -(\rho_{he}^{ji})^T = (\rho_{eh}^{ji})^\dagger.
\label{rho}
\eeqa
The symbol $\otimes$ denotes a direct product, so that each of the above expressions is a matrix over orbital space, e.g. $\rho_{es,es'}^{ij} = \langle c_{js'}^\dagger c_{is}\rangle$. Both $\rho_{ee}$ and $\rho_{hh}$ are Hermitian, and $\rho_{eh}$ and $\rho_{he}$ are antisymmetric as a result of anticommutation relations.

We can define Nambu spinors so that $e,h$ become additional quantum numbers: $\check{c}^\dagger_\bi = (c^\dagger_{\bi}, c_{\bi})$ and $\check{c}_\bi = (c_{\bi}, c^\dagger_{\bi})$. [If we only have spin as orbital quantum numbers, we obtain 4-component spinors $\check{c}^\dagger_\bi = (c^\dagger_{i\uparrow}, c^\dagger_{i\downarrow}, c_{i\uparrow}, c_{i\downarrow})$.] The corresponding Nambu density matrix reads
\beq
\check{\rho}^{ij} = \langle \check{c}_\bj^\dagger\otimes \check{c}_\bi\rangle = \left( \begin{array}{cc}
    \rho_{ee}^{ij} & \rho_{eh}^{ij} \\
    \rho_{he}^{ij} & \rho_{hh}^{ij}
\end{array} \right) =
\left( \begin{array}{cc}
    \rho_{ee}^{ij} & \rho_{eh}^{ij} \\
    \rho_{he}^{ij} & \mathbbm{1}\delta_{ij}-(\rho_{ee}^{ji})^T
\end{array} \right).
\eeq

The mean-field Hamiltonian operator can be written in terms of Nambu spinors as
\beqa
\label{HNambuSQ}
H &=& \frac{1}{2}\sum_{\bi\bj}\check{c}^\dagger_\bi
(\check{H}_0^{\bi\bj} + \check{\Sigma}^{\bi\bj})
\check{c}_\bj, \\
\check{H}_0^{ij} &=&\left( \begin{array}{cc}
    H_0^{ij} & 0 \\
    0 & -{H_0^{ji}}^T
\end{array} \right), \\
\check{\Sigma}^{ij} &=&\left( \begin{array}{cc}
    \delta_{ij}\Sigma_H^{ii} + \Sigma_F^{ij} & \Sigma_B^{ij} \\
    {\Sigma_B^{ji}}^\dagger & -(\delta_{ij}\Sigma_H^{ii} + \Sigma_F^{ji})^T
\end{array} \right).
\label{HNambu}
\eeqa
Note that the $1/2$ in Eq. \eqref{meanfield} is accounted for by the $1/2$ in Eq. \eqref{HNambuSQ}.

We would now like to express $\check\Sigma$ more compactly in terms of $\check\rho$ using Eqs. \eqref{Sigma}. The Nambu density matrix $\check\rho$ does not have the same symmetries as the Nambu Hamiltonian $\check{H}$ or $\check\Sigma$. However, the following object does
\beq
\tilde{\rho}^{ij} = \left( \begin{array}{cc}
    \rho_{ee}^{ij} & \rho_{eh}^{ij} \\
    \rho_{he}^{ij} & -(\rho_{ee}^{ji})^T
\end{array} \right) =
\check{\rho}^{ij}-
\delta_{ij}\left( \begin{array}{cc}
    0 & 0 \\
    0 & \mathbbm{1}
\end{array} \right).
\eeq
This modified density matrix allows us to write the Nambu mean-field Hamiltonian in a compact form
\beq
\label{SigmaNambu}
\check{\Sigma}^{ij} =  \frac{1}{2}\delta_{ij}\tau_z\sum_l v^{il} \Tr(\tau_z \tilde{\rho}^{ll})-v^{ij}\tau_z\tilde{\rho}^{ij}\tau_z.
\eeq
The first term corresponds to the Hartree mean field, and the second one to both the Fock and the Bogoliubov mean fields. Note that in the spinless case the $i=j$ Fock term perfectly cancels the self-interaction $i=j=l$ Hartree term, as expected from Eq. \eqref{Hint0}.

If we specialize to the (spinful) Hubbard model, where the interaction reads $U\sum_i n_{i\uparrow}n_{i\downarrow}$ (here $n_{i\sigma}=c^\dagger_{i\sigma}c_{i\sigma}$ is the number operator), we need to set $v^{ij} = \delta_{ij}U$, which yields
\beq
\check{\Sigma}^{ij} =  U\delta_{ij}\left(\frac{1}{2}\Tr(\tau_z \tilde{\rho}^{ii})\tau_z - \tau_z\tilde{\rho}^{ii}\tau_z\right).
\label{HFN}
\eeq

\section{Generalized gap equations without SOC}
\label{ap:HFBnosoc}

Let us consider $H_0$ for an infinite nanowire without SOC but with Zeeman field $V_Z$,
\beq
H_0(k)=\sum_{ks}c^\dagger_{ks}\left[(\epsilon_k-\mu)\delta_{ss'} +V_Z\sigma_{ss'}\right]c_{ks'}.
\eeq
where $k$ is a wavevector, not a site index, and $\epsilon_k$ is the spin-independent dispersion relation.

We wish to spell out Eq. \eqref{HFN} for this case, to obtain a generalized version of the traditional self-consistent BCS gap equation, see Eq. \eqref{gapeq0main}, but now using the full Hartree-Fock-Bogoliubov mean field theory. We rewrite Eq. \eqref{HFN} for $i=j$ as
\beq
\check{\Sigma}^{ii} = (\delta V_Z\sigma_z-\delta\mu \sigma_0)\tau_z+\Delta\sigma_y\tau_y.
\eeq
This is a re-parametrization of the $4\times 4$  matrix  $\check\Sigma^{ii}$ consistent with the symmetries of Eq. \eqref{rho}, in which $\Delta$ is assumed to be real. Using this form, we can write the off-diagonal $\Sigma^{ii}_{eh}$ block of Eq. \eqref{HFN} as $-i\sigma_y\Delta = U \rho_{eh}^{ii}$, so in particular
\beq
\Delta = U\rho_{e\downarrow, h\uparrow}^{ii}.
\label{gapsc}
\eeq
For the $H_0(k)$ at hand, we have
\beq
\rho_{e\downarrow, h\uparrow}^{ii} = \frac{1}{N}\sum_{k}\left[ u_{k+} v_{k+}^* f(E_{k+})+u_{k-}v_{k-}^* f(E_{k-})\right],
\label{rhoH1}
\eeq
where we have used $\check{\rho}^{ij} = \frac{1}{N}\sum_k \langle \check{c}^\dagger_k\otimes \check{c}_k\rangle e^{k(x_j-x_i)}$. Here $f(\epsilon) = 1/(e^{\epsilon/k_BT}+1)$ is the Fermi-Dirac distribution, $k$ is summed over all $N$ wavevectors in the Brillouin zone ($N$ is also the total number of sites in the system). Note that $\rho_{e\uparrow, h\downarrow}^{ii} = -\rho_{e\downarrow, h\uparrow}^{ii}$.

In Eq. \eqref{rhoH1}, $E_{k\pm}$ and $(u_{k\pm}, v_{k\pm})$ are, respectively, the eigenvalues and (normalized) eigenvectors of the $2\times 2$ $e\downarrow/h\uparrow$ block of $\check{H}(k) = \check{H}_0(k)+\check{\Sigma}$, namely
\beq
\label{Hkblock}
\check{H}(k)^{e\downarrow/h\uparrow} = \left( \begin{array}{cc}
    \tilde{\epsilon}_k - \tilde{V}_Z & \Delta \\
    \Delta & -\tilde{\epsilon}_k - \tilde{V}_Z
\end{array} \right).
\eeq
The dressed normal eigenenergies and Zeeman fields are
\beqa
\tilde{\epsilon}_k &=& \epsilon_k-\mu-\delta\mu,\\
\tilde{V}_Z &=& V_Z+\delta V_Z,
\eeqa
which contain the $U$-induced shifts to the Fermi energy $\delta\mu$ and Zeeman field $\delta V_Z$, respectively.
We have $E_{k\pm} = \tilde{V}_Z\pm\sqrt{\Delta^2+\tilde{\epsilon}_k^2}$, and $(u_{k\pm}, v_{k\pm})$ are also straightforward. Simplifying Eq. \eqref{gapsc} with Eq. \eqref{rhoH1} we obtain the self-consistent gap equation
\beq
\Delta = -\frac{U}{N}\sum_k\frac{\Delta}{4\tilde{E}_k}\left[\tanh\!\left(\frac{\tilde{E}_k+\tilde{V}_Z}{2k_BT}\right)\!+\tanh\!\left(\frac{\tilde{E}_k-\tilde{V}_Z}{2k_BT}\right)\right],
\label{gapeq}
\eeq
where $\tilde{E}_k = \sqrt{\Delta^2+\tilde{\epsilon}_k^2}$.

Ignoring $V_Z$, $\delta V_Z$ and $\delta\mu$, this reduces to the traditional gap equation often found in textbooks~\cite{Tinkham:04} that is usually derived for an infinite system without Zeeman and SOC,
\beq
\label{gapeq0}
\Delta = - \frac{U}{N}\sum_k\frac{\Delta}{2E_k}\tanh\left(\frac{E_k}{2k_BT}\right),
\eeq
where $E_k = \sqrt{\Delta^2+\epsilon_k^2}$. Note that this has a nonzero solution only if $U<0$ (attractive interaction).

In addition to the generalized Eq. \eqref{gapeq}, we have  implicit, self-consistent conditions for $\delta V_Z$ and $\delta\mu$ as well. These can be obtained by considering the diagonal elements of $\check{\Sigma}^{ii}$
\beq
\delta V_Z\sigma_z - \delta\mu\sigma_0 = U\left(\begin{array}{cc}
    \rho_{ee}^{i\downarrow,i\downarrow} & 0 \\
    0 & \rho_{ee}^{i\uparrow,i\uparrow}
\end{array}\right),
\eeq
where
\beq
\rho_{e\downarrow,e\downarrow}^{ii} = \frac{1}{N}\sum_{k}\left[|u_{k+}|^2 f(E_{k+})+|u_{k-}|^2 f(E_{k-})\right],
\eeq
and a similar $\rho_{e\uparrow,e\uparrow}^{ii}$ with flipped sign of $\Delta$ and $V_Z$. The complete set of self-consistent equations is then
\begin{widetext}
\beqa
\Delta &=& -\frac{U}{4N}\sum_k\frac{\Delta}{\tilde{E}_k}\left[\tanh\!\left(\frac{\tilde{E}_k+\tilde{V}_Z}{2k_BT}\right)\!+\tanh\!\left(\frac{\tilde{E}_k-\tilde{V}_Z}{2k_BT}\right)\right],\\
\delta\mu &=& -\frac{U}{2}+\frac{U}{4N}\sum_k\frac{\tilde{\epsilon}_k}{\tilde{E}_k}\left[\tanh\!\left(\frac{\tilde{E}_k+\tilde{V}_Z}{2k_BT}\right)\!+\tanh\!\left(\frac{\tilde{E}_k-\tilde{V}_Z}{2k_BT}\right)\right],\\
\delta V_Z &=&\frac{U}{2N}\sum_k\frac{\sinh(\tilde{V}_Z/k_BT)}{\cosh(\tilde{V}_Z/k_BT)+\cosh(\tilde{E}_k/k_BT)}.
\eeqa
\end{widetext}
If we impose $\delta\mu = \delta V_Z = V_Z = 0$ we recover the conventional gap equation, Eq. \eqref{gapeq0}, or Eq. \eqref{gapeq0main} in the main text. We emphasize that the above generalization of the gap equation is made possible by neglecting SOC, so that the Hamiltonian $\check{H}(k)$ can be decomposed into two decoupled blocks like Eq. \eqref{Hkblock}.

It is important to note that $\Delta$ in the gap equation is only the gap of the system if $V_Z=0$. The actual gap is denoted as $\Omega$ in the main text and corresponds to the minimal eigenvalue of $\check{H}(k)$. The gap $\Omega$ without SOC can be reduced to
\beq
\label{Omega}
\Omega = \max(0, |\Delta| - |V_Z|).
\eeq
This equation corresponds to the dashed line in Fig. \ref{fig:2}(e).

We can solve the conventional gap equation \eqref{gapeq0} analytically in the zero-temperature small-gap limit. By linearizing the dispersion as $\epsilon_k = \pm v_F (k\mp k_F)$ where $v_F$ and $k_F$ are Fermi velocity and wavevector, respectively, and turning the sum $\frac{1}{N}\sum_k$ into an integral over $\epsilon_k$, $\frac{a_0}{2\pi v_F}\int_{-\epsilon_0}^{\epsilon_0} d\epsilon_k$, where $\epsilon_0$ is a phenomenological cutoff, Eq. \eqref{gapeq0} becomes
\beq
\Delta = \frac{-2Ua_0\Delta}{2\pi v_F}\int_{-\epsilon_0}^{\epsilon_0} d\epsilon \frac{1}{\sqrt{\Delta^2+\epsilon^2}}.
\eeq
This equation only has a non-zero solution $\Delta$ for $U<0$. In the limit $\Delta\ll \epsilon_0$, it simplifies to
\beq
\label{gapeqsmall}
\Delta=\Delta\frac{2|U|a_0}{\pi v_F}\ln\left(\frac{2\epsilon_0}{\Delta}\right).
\eeq
For $U<0$ the pairing potential then becomes
\beq
\label{gapsol0}
\Delta = 2\epsilon_0 \exp\left(-\frac{\pi \hbar v_F}{2|U|a_0}\right),
\eeq
where we have restored $\hbar$ for clarity. This is the textbook BCS solution~\cite{Tinkham:04}.

This solution is also valid at finite $|V_Z|\leq \Delta$, since the $\tanh$ factors in Eq. \eqref{gapeq} remain equal to 1 throughout the integration interval. For $|V_Z|>\Delta$ the gap $\Omega$ closes, see Eq. \eqref{Omega}. The pairing, however, remains finite. The generalization of Eq. \eqref{gapeqsmall} for $|V_Z|>\Delta$ becomes
\beqa
\label{gapVzbig}
\Delta&=&\Delta\frac{2|U|a_0}{\pi v_F}\left[\ln\left(\frac{2\epsilon_0}{\Delta}\right)\right.\\
&&\left.+\frac{1}{2}\ln\left(\frac{2V_Z(V_Z-\sqrt{V_Z^2-\Delta^2})-\Delta^2}{\Delta^2}\right)\right].\nonumber
\eeqa
Its solution can be formulated in terms of the pairing at $V_Z=0$, Eq. \eqref{gapsol0}. Renaming this pairing as $\Delta_0 = \Delta(V_Z=0)$, the solution of Eq. \eqref{gapVzbig} can be expressed as
\beq
\Delta = \frac{\sqrt{\Delta_0(2|V_Z|-\Delta_0)}}{2\epsilon_0},
\eeq
subject to the condition that $|V_Z|>\Delta$, which implies in particular $|V_Z|>\Delta_0/2$, so the root is real.

\end{document}